\documentclass[showpacs,amsmath,amssymb,aps,twocolumn,superscriptaddress,prx]{revtex4-2}
\usepackage{mathtools}
\usepackage{algorithm}
\usepackage{algpseudocode}
\usepackage{multirow}
\usepackage[pdftex]{graphicx} \graphicspath{{}}
\usepackage{grffile} 
\usepackage{comment}
\usepackage{makecell}
\usepackage{braket}
\usepackage{placeins}
\usepackage{diagbox}
\usepackage{bbding}
\usepackage{ulem}
\usepackage{float}
\usepackage{afterpage}
\usepackage{dcolumn}
\usepackage{bm}
\usepackage[pdftex,colorlinks=true]{hyperref}
\usepackage[dvipsnames]{xcolor}
\definecolor{darkblue}{HTML}{1B436C}
\definecolor{darkred}{HTML}{94080A}
\hypersetup{
	colorlinks=true,
    urlcolor=darkred,
    citecolor=darkblue,
    linkcolor=darkblue
}
\definecolor{ForestGreen}{RGB}{34, 139, 34}

\newcommand{\affA}{State Key Laboratory of Artificial Microstructure and Mesoscopic
Physics, School of Physics, Peking University, Beijing 100871, China}
\newcommand{\affB}{Max Planck Institute for the Physics of Complex Systems, N\"othnitzer Str.~38, 01187 Dresden, Germany}
\newcommand{\affC}{Cavendish Laboratory, University of Cambridge, Cambridge CB3 0HE, United Kingdom}
\newcommand{\affD}{School of Physical Science and Technology, ShanghaiTech University, Shanghai 201210, China}
\newcommand{\affE}{Technical University of Munich, TUM School of Natural Sciences,
Physics Department, 85748 Garching, Germany}
\newcommand{\affF}{Munich Center for Quantum Science and Technology (MCQST), Schellingstr. 4, 80799 M{\"u}nchen, Germany}
\newcommand{\affG}{Blackett Laboratory, Imperial College London, London SW7 2AZ, United Kingdom}

\begin{document}

\title{Floquet Superheating
 }

\author{Yang Hou}
 \affiliation{\affA}
  \affiliation{\affB}
\author{Andrea Pizzi}
 \affiliation{\affC}
 \author{Huike Jin}
 \affiliation{\affD}
	\author{Johannes Knolle}
\affiliation{\affE}
\affiliation{\affF}
\affiliation{\affG}
	\author{Roderich Moessner}
 \affiliation{\affB}
 	\author{Hongzheng Zhao}
	\email{hzhao@pku.edu.cn}
 \affiliation{\affA}
	\date{\today}

\begin{abstract}
Periodically driven many-body systems generally heat towards a featureless `infinite-temperature' state. As an alternative to uniform heating in a clean system, 
here we establish a Floquet superheating regime, where fast  heating nucleates  at ``hot spots" generated by rare fluctuations in the local energy with respect to an appropriate effective Hamiltonian. Striking
macroscopic consequences include exceptionally long-lived prethermalization and non-ergodic bimodal distributions of macroscopic observables. Superheating is predicated on a heating rate depending strongly on the local fluctuation; in our example, this is supplied by a sharp state-selective spin echo, where the energy absorption is strongly suppressed for low-energy states, while thermal fluctuations open up excessive heating channels. A simple phenomenological theory is developed to show the existence of a critical droplet size, which incorporates heating by the driving field as well as the heat current out of the droplet. Our results shine light on a new heating mechanism and suggest new routes towards stabilizing
non-equilibrium phases of matter in driven systems.
\end{abstract}
	\maketitle
\let\oldaddcontentsline\addcontentsline
\renewcommand{\addcontentsline}[3]{}

\textit{Introduction.---} 
A closed Hamiltonian many-body system with a time-dependent drive tends to absorb energy, eventually reaching an equilibrium state that is entirely featureless~\cite{Lazarides2014Equilibrium}.
Increasing the driving frequency can exponentially suppress heating, yielding a transient but long-lived prethermal regime~\cite{bukov2015prethermal,abanin2015exponentially,kuwahara2016floquet}, which has been exploited to realize various non-equilibrium phases of matter~\cite{kitagawa2010topological,khemani2016phase,else2016floquet,yao2017discrete,russomanno2017floquet,else2020long,mcginley2022absolutely}. In clean systems, this heating process occurs, apparently typically,
with a rate that depends on macroscopic properties~\cite{ikeda2021fermi,mori2022heating}, e.g., the system’s energy, while microscopic fluctuations appear not to play an essential role.

This can change when the system’s behavior is dominated not by small fluctuations around equilibrium, but by the tails of the probability distribution—the rare events~\cite{krapivsky2010kinetic}. Famous examples include superheating, in which water is brought above its boiling point without boiling, and supercooling, in which it is cooled below its freezing point without freezing. 
The ability of the system to remain in the “wrong” metastable phase is tied to first-order phase transitions~\cite{landau1937theory,rikvold1994metastable,lagnese2024detecting}: the transformation to the stable phase is not automatic but instead must proceed through the rare nucleation of a droplet and its subsequent growth. Therefore, rare microscopic events can induce striking macroscopic consequences~\cite{agarwal2017rare,carollo2018making}.

A natural question is whether Floquet heating can be initiated by rare events—in fundamental distinction to the familiar homogeneous process. 
This is challenging to address
because rare events are, by their very nature, difficult to access. They tend not to leave noticeable macroscopic signatures until long times. Moreover, distinguishing rare events from homogeneous heating—which can occur rapidly and dominate the dynamics—requires careful control over the system and driving parameters.

\begin{figure*}[t]
    \centering
    \includegraphics[width=\linewidth]{ 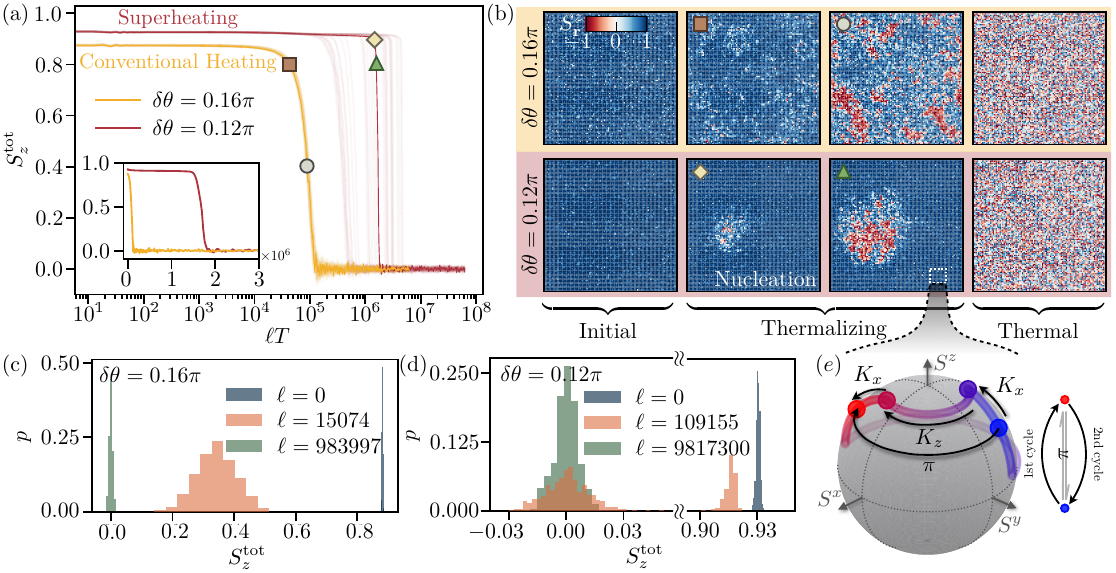}
    \caption{(a) Dynamics of conventional heating (yellow traces) and superheating (red traces) as reflected in the total $z-$magnetization. Each ensemble depicts 32 individual realizations, which stay close to each other conventionally, but not for superheating. We use  $L=100$, $J=0.5$, $g=0.17$ for numerical simulations.
(b) Snapshots of the $z$-component spin configurations for conventional heating(up) and superheating (down). Floquet superheating occurs via the nucleation process of a `hot spot', which quickly expands and destabilizes the system. The markers correspond to those in (a). The color denotes the value of the $z-$magnetization, such that blue and red colors correspond to spins pointing along the positive and negative $z$-directions, respectively. 
    (c) (d) Probability distribution of the total magnetization. A bimodal distribution appears during the superheating for low-energy initial states, while a Gaussian unimodal distribution appears for conventional heating with higher energy initial states. 512 independent realizations are used here.
    (e) 
    Schematic for the state-selective spin echo. Ising interaction effectively echoes out the transverse field, hence suppressing heating of the low-energy background.}
    \label{fig.dynamics}
\end{figure*}

Here, we provide an affirmative answer and show that \textit{Floquet superheating} can occur, {by identifying a setting where heating takes place uniformly, but very slowly, until fast heating} is triggered locally by rare and isolated events. Starting from a low-energy initial state with weak spatial fluctuations, the system can thus remain stable for exceedingly long times. The triggering takes place at “hot spots” with higher local energy density [see Fig.~\ref{fig.dynamics}(b)], which spontaneously arise not because of any quenched disorder, but due to the spontaneous fluctuations of the energy density always present across the system. {In a slight abuse of terminology, we coin the term Floquet superheating to refer to the combination of a slowly-heating and long-lived prethermal regime, and the runaway heating process which terminates it.}

Concretely, we present a Floquet-driven 2D Ising system  for an explicit demonstration. For numerical efficiency, we perform large-scale, long-time simulations of classical spin dynamics~\cite{mori2018floquet,howell2019asymptotic,pizzi2021classical,ye2021floquet,guo2025emergent,kim2025confined}, although our conclusions should equally apply to quantum systems. We show that rare nucleation events occur at times that vary greatly across different initial-state realizations. Consequently, for ensemble average over realizations the system develops a non-ergodic bimodal distribution of macroscopic quantities, such as total magnetization, which is sharply distinct from homogeneous heating, where a single trajectory typically follows that of a thermal ensemble.

We map out a `phase diagram' and show that Floquet superheating arises only in certain intermediate-frequency regimes, underpinned by an emergent state-selective spin echo (SSSE). Crucially, unlike a conventional spin echo, which relies on an external field~\cite{choi2020robust,fu2024engineeringhierarchicalsymmetries}, here intrinsic many-body interactions effectively echo out heating channels at low energies. The spontaneous emergence of a higher-energy droplet breaks the SSSE, thereby locally opening excessive heating channels through the droplet’s bulk. A simple phenomenological rate equation captures this droplet dynamics. Importantly, as the surrounding low-energy region can autonomously cool the droplet, expansion only happens when the droplet exceeds a critical size, such that energy absorption inside the bulk dominates dissipation through the surface. The critical droplet thus sets a barrier that the system needs to cross before the rapid heating effects of Floquet thermalization, with a nucleation probability that is exponentially suppressed in the droplet size.

\textit{Model.---} We consider $L^2$ classical spins on a square lattice with a time periodic Hamiltonian $H(t){=}H(t{+}T)$ with driving period $T$, 
\begin{align}\label{eq. Hamiltonian}
    H(t){=}
    \begin{cases}
     H_1{\equiv} g \sum_{\boldsymbol{r}} S_{\boldsymbol{r}}^x\quad\text{for}\ t\in \left[0,\frac{T}{4} \right)\cup\left[\frac{3T}{4},T \right)\\[8pt]
    H_2{\equiv} - \frac{J}{4}\sum_{{\boldsymbol{r}}{,}\boldsymbol{\delta}}S_{\boldsymbol{r}}^zS_
    {{\boldsymbol{r}}{+}\boldsymbol{\delta}}^z\quad \text{for}\ t\in \left[\frac{T}{4},\frac{3T}{4}\right)
    \end{cases}\ ,
\end{align}
where $\boldsymbol{r}$ runs through the lattice sites and $\boldsymbol{\delta}{=}{\pm} \hat{x},{\pm}\hat{y}$ through the lattice vectors. We consider periodic boundary conditions. The normalized spin variable $\boldsymbol{S}_{\boldsymbol{r}}$ on site ${\boldsymbol{r}}$ can be parametrized by an azimuthal angle $\varphi_{\boldsymbol{r}}$ and a polar angle $\theta_{\boldsymbol{r}}$ as $S_{\boldsymbol{r}}^x{=}\sin{\theta_{\boldsymbol{r}}}\cos{\varphi_{\boldsymbol{r}}}$, $S_{\boldsymbol{r}}^y=\sin{\theta_{\boldsymbol{r}}}\sin{\varphi_{\boldsymbol{r}}}$, $S_{\boldsymbol{r}}^z=\cos{\theta_{\boldsymbol{r}}}$. The spins 
satisfy the Poisson bracket $\{S_{\boldsymbol{r}}^\mu,S_{{\boldsymbol{r}}'}^\nu\}=\delta_{{\boldsymbol{r}}{\boldsymbol{r}}'}\epsilon^{\mu\nu\rho}S_{\boldsymbol{r}}^{\rho}$, with the antisymmetric tensor $\epsilon^{\mu\nu\rho}$, and 
undergo Hamiltonian dynamics $\dot{S}_{\boldsymbol{r}}^\mu {=} \{S_{\boldsymbol{r}}^\mu, H(t)\}$. Here, $H_1$ consists of a magnetic field and induces a uniform spin precession about the $x$ axis with frequency $g$; $H_2$ comprises instead Ising interactions of strength $J$, inducing a nonlinear rotation around the $z$ axis that depends on the average effective field of the neighboring spins~\footnote{The dynamics for both $H_1$ and $H_2$ can be integrated analytically, enabling highly efficient large-scale and long-time numerical simulations~\cite{howell2019asymptotic}.}.

{\it Superheating Phenomenon.---}
The system is initialized with $\varphi_{\boldsymbol{r}}$ randomly sampled within $(0,2\pi)$ and  $\theta_{\boldsymbol{r}}$ drawn from a Gaussian distribution with mean zero and standard deviation $\delta\theta$. {For high-frequency drives, the effective Hamiltonian $H_{\mathrm{eff}}{=}(H_1{+}H_2)/2$ is a quasi-conserved quantity; $H_{\mathrm{eff}}$ can be used to quantify the system's energy, and its initial value is controlled by $\delta\theta$.} For small $g$, the low-energy state is ferromagnetic with a high magnetization density along $z$, see Fig.~\ref{fig.dynamics}(b).

We first focus on the conventional uniform heating.
In Fig.~\ref{fig.dynamics}(a), we present the dynamics of the total $z$ magnetization density ($S_z^{\mathrm{tot}}$), initialized from two different ensembles of initial energies, each consisting of 32 individual realizations. {We use $\ell$ to label the number of driving cycles in Fig.~\ref{fig.dynamics}.} 
For $\delta\theta{=}0.16\pi$ (yellow), different trajectories lead to approximately the same evolution, which saturates at a prethermal plateau before the onset of heating around $t{\sim} 10^4$. {At long times, both the magnetization and the energy decay to zero, corresponding to the infinite-temperature expectation value.}
In Fig.~\ref{fig.dynamics}(d), we also depict the probability distribution of $S_z^{\mathrm{tot}}$ at different times, using 512 independent realizations, where a Gaussian-like unimodal distribution is observed throughout the entire evolution.
In fact, as shown in Fig.~\ref{fig.dynamics}(b), the spin configuration for a given realization at different times (cf.~markers in Fig.~\ref{fig.dynamics}(a) and (b)) confirms that heating occurs homogeneously in space, as reported in other studies of Floquet heating~\cite{rubio2020floquet,ye2020emergent,pizzi2021classical,ikeda2021fermi}.

For weaker spatial disorder $\delta\theta{=}0.12\pi$, however,
Floquet superheating is triggered, and rapid heating occurs at strikingly distinct times in different trajectories; see Fig.~\ref{fig.dynamics}(a). 
On average, this prethermal lifetime ($\langle\tau\rangle{\sim} 10^6$) is two orders of magnitude longer than the previous case (yellow, $\langle\tau\rangle{\sim} 10^4$), with notable fluctuations comparable to its mean, {a typical sign of non-ergodicity}. 
This also leads to a bimodal probability distribution during the heating process with a large relative standard deviation $r{=}{\sqrt{\langle\tau^2\rangle{-}\langle\tau\rangle^2}}/{\langle\tau\rangle} {\sim} \mathcal{O}(1)$. For example, as shown in Fig.~\ref{fig.dynamics}(c), two peaks of approximately equal weight appear around $\ell{\approx}10^5$ (red) {driving cycles}, where the total magnetization concentrates around either $S_{z}^{\mathrm{tot}}{=}0$ or 0.91. Only a negligible fraction of trajectories exhibit a value of $S_z^{\mathrm{tot}}$ between these two extreme values. 

To understand this behavior, we again plot the spin configurations at different times in Fig.~\ref{fig.dynamics}(b). In particular, 
at the onset of heating (diamond marker), a high-energy droplet spontaneously nucleates out of the low-energy background, a characteristic signature of Floquet superheating. Energy absorption concentrates on this droplet, which rapidly grows and heats up the entire system within a short time $\tau_{\text{grow}}$ {for the current system size}.  
The formation of the droplet is rare and typically occurs after an exceptionally long time $\tau_{\mathrm{nuc}}$, which determines the prethermal lifetime. As $\tau_{\text{grow}}{\ll} \tau_{\mathrm{nuc}}$, in Fig.~\ref{fig.dynamics}(a) we see an abrupt drop of the total magnetization during the evolution (see also the inset with a linear time axis). We highlight that the time and location of nucleation fluctuate significantly among trajectories, despite all starting from macroscopically similar initial states. The large fluctuation in 
$\tau_{\mathrm{nuc}}$ and the fact $\tau_{\mathrm{nuc}}{\gg} \tau_{\text{grow}}$ result in the bimodal distribution in Fig.~\ref{fig.dynamics}(c). 

{For larger $L$, $\tau_{\text{grow}}$ increases, whereas the typical nucleation time $\langle\tau\rangle$ is independent of $L$ as it originates from local spontaneous fluctuations. Therefore, for $L{\to}\infty$, a finite density of droplets can coexist in a single realization. However, for a sufficiently large region involving $10^4$ spins that we can numerically simulate, in most realizations we only observe one droplet nucleation event (see one example with two droplets in Sec.~1.2  in the Supplementary Material (SM)~\cite{SM}).

Floquet superheating appears for low-energy initial states across a wide range of driving frequencies $\omega$. To better quantify this, the average prethermal lifetime $\langle \tau\rangle$ and the relative standard deviation $r$ serve as suitable measures~\footnote{We extract $\tau$ as the time when $S_{z}^{\mathrm{tot}}$ drops below 80 percent of its initial value}.
We fix $\delta\theta{=}0.12\pi$ and plot $\langle \tau\rangle$ and $r$ in Fig.~\ref{fig.phase}(a) and (b) (red lines), respectively,
for various frequencies. The high-frequency regime, $\omega{\gtrsim}1.25$, yields the expected exponential scaling $\langle \tau \rangle {\sim} e^{\mathcal{O}(\omega)}$ and small fluctuations $r$.
By contrast, a non-monotonic dependence appears for intermediate $\omega$, with $\langle\tau\rangle$ peaking around $\omega{\approx}1$. Remarkably, a slight change in $\omega$ induces a variation of $\langle\tau\rangle$ by several orders of magnitude, e.g., $\langle\tau\rangle$ rises from $10^5$ to $10^7$ as $\omega$ varies from 0.95 to 1. The concomitant standard deviation $r$ also takes the value of $\mathcal{O}(1)$~\footnote{We use 512 realizations for most of the data points in Fig.~\ref{fig.phase}. For parameters with a large value of $r$, 2048 realizations are used to ensure a better convergence.}, in accordance with the bimodal distribution in Fig.~\ref{fig.dynamics}(c).

We map out a `phase diagram' in Fig.~\ref{fig.phase}(c), where the color denotes the averaged prethermal lifetime $\langle\tau\rangle$ on a log scale. Clearly, Floquet superheating appears in the dark region around $\omega{=}1$ for sufficiently low-energy initial states. Yet, in the high-frequency regime, heating is always exponentially suppressed with a weaker dependence on the strength of the initial spatial randomness $\delta \theta$. 
The gray squares correspond to the critical frequencies, which clearly separate parameter regimes that lead to Floquet superheating from the homogeneous heating.

\begin{figure}[t]
    \centering
    \includegraphics[width=\linewidth]{ 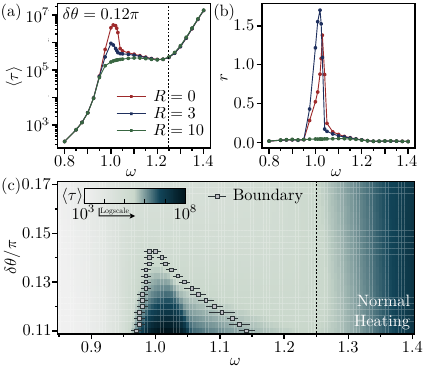}
    \caption{(a) Mean value $\langle\tau\rangle$ of the prethermal lifetime. For large $\omega$, an exponential scaling appears. For intermediate frequencies, instead, $\langle\tau\rangle$ peaks around $\omega{\approx}1$ where superheating occurs. Seeding a droplet of linear size $R$ generally speeds up heating. (b) The relative standard deviation of $\tau$ exhibits a notable peak for $R{=}0,3$, indicating that superheating occurs due to rare events. For $R{=}10$, the initially seeded droplet induces fast heating deterministically. We use $\delta\theta{=}0.12\pi$ with $J{=}0.5$, $g{=}0.17$, and $L{=}100$ for numerical simulations.
    (c) $\langle\tau\rangle$  on a log scale for $L{=}50$. The dark region around $\omega{\approx}1$ features Floquet superheating. The gray squares correspond to the critical frequencies, obtained by averaging over three values of $\omega$ where $r$ exceeds the threshold values $0.3,0.35,0.4$, with error bars denoting their standard deviation. 
    }
    \label{fig.phase}
\end{figure}

\textit{State-selective Spin Echo.--} We next identify and characterize the mechanism underpinning Floquet superheating: an SSSE suppresses the homogeneous heating rate such that the rare nucleation events become visible.

At the mean-field level, we approximate the Ising Hamiltonian as $\bar{H}_2{=}{-}2J\bar{S}^z\sum_{\mathbf{r}}S_{\mathbf{r}}^z,
$
with $\bar{S}^z$ the averaged $z$-component of the spins. For each spin and after half of a driving period $T/2$, this mean-field generates a rotation around the $z$ axis by $\phi{=}{2J}\bar{S}^z\pi/{\omega}$, with $\bar{S}^z\approx1$ for sufficiently ordered ferromagnetic states. See schematic in Fig.~\ref{fig.dynamics}(e) where $K_z$ denotes this rotation. For 
$\omega={2J\bar{S}^z}/(2n{+}1)$ with integer $n$, the rotation angle $\phi$ becomes an odd multiple of $\pi$. $K_z$ effectively echoes out the rotations along the transverse field ($K_x$ in Fig.~\ref{fig.dynamics}(e)) every $2T$. The system consequently returns to its origin, following the closed orbit shown in Fig.~\ref{fig.dynamics}(e). {A refined description of SSSE is shown in Sec.~4.1 in the SM~\cite{SM}.}
\begin{figure}
    \centering
\includegraphics[width=\linewidth]{ 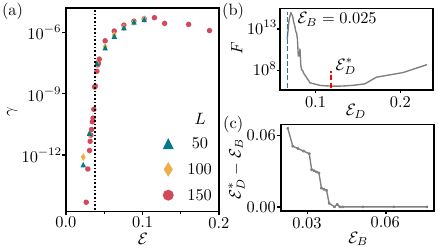}
    \caption{ (a) Heating rate $\gamma$ versus initial energy (measured w.r.t the ground state energy) {for parameters corresponding to SSSE. For sufficiently low-energy initial states, $\gamma$ is greatly reduced due to SSSE and a sharp drop is observed. For the ground state, heating is strictly forbidden. 
    }(b) Phenomenological result for $F(\mathcal{E}_D)$ in Eq.~\eqref{eq.probRc} for a low background energy $\mathcal{E}_B$. A global minimum $\mathcal{E}_D^*$ different from $\mathcal{E}_B$ exists, suggesting that
    nucleation events can occur spontaneously. (c) Spontaneous
    nucleation only occurs when the difference between  $\mathcal{E}_D^*$ and $\mathcal{E}_B$ is finite. For large $\mathcal{E}_B$, this is not allowed.}
    \label{fig.rate}
\end{figure}

{SSSE is fundamentally different from the conventional echoing mechanism, where normally an external field echoes out unwanted heating channels~\cite{choi2020robust,FleckensteinPrethermalization2021}, leading to, e.g., emergent symmetries in the time evolution operator~\cite{khemani2016phase,else2017prethermal}. In contrast, SSSE relies on the intrinsic many-body interaction and only stabilizes special low-energy states, and hence the suppression of the heating rate $\gamma$ exhibits a strong initial state dependence; see detailed comparison in Sec.~2 in the SM~\cite{SM}.} In Fig.~\ref{fig.rate}(a) we consider $\omega{=}1$ (i.e., $n{=}0$) and various initial energy densities (i.e., $\delta \theta$) and plot $\gamma$, which is extracted by tracking the slow increase in system's energy {$H_{\mathrm{eff}}$} before the first nucleation event, see details in Sec.~3.2 in the SM~\cite{SM}}. {Note, in the following discussion we measure the energy density w.r.t. the ground state, ${H}_{\mathrm{eff}}^{\mathrm{GS}}/L^2{\approx}{-}0.257$.}
Clearly, $\gamma$ is strongly suppressed for sufficiently low-energy initial states.
Increasing the initial state energy, and hence spatial fluctuations, perturbs the aforementioned closed orbits and increases $\gamma$, with a remarkably sharp drop near $\mathcal{E}{\approx}0.03$ (black dotted line in Fig.~\ref{fig.rate}(a)).

Next, we initialize a local `seed' droplet of high energy in the initial state for a controlled study of droplet dynamics. Concretely, we randomize $R^2$ spins in a square region in the initial state and show $\langle \tau\rangle$ and $r$ in 
Fig.~\ref{fig.phase}. The presence of the droplet generally shortens $\langle \tau\rangle$ (cf. the blue and red data in panel (a)). Yet, as shown in panel (b), only when the droplet is sufficiently large (green) does the relative standard deviation $r$ reduce to zero at $\omega\approx1$. This observation thus strongly suggests that only when the droplet exceeds a critical size, fast heating can be triggered. The critical droplet size can also be extracted numerically, see Sec.~3.1 in the SM~\cite{SM}.

{\it Droplet Theory.---}
{Next, we develop a droplet theory to justify the existence of a critical droplet size.}
In standard nucleation theory, the competition between the droplet’s bulk energy and surface tension determines whether the droplet expands or not~\cite{rikvold1994metastable}. Here, we present a dynamical version of this picture for our Floquet setting, where it is the competition between the {\it rates} of heating in the bulk and energy diffusion through the surface that determines the critical droplet size.

Consider a spherical droplet of radius $R$ with a uniform interior energy density $\mathcal{E}_D$ on top of a background with a low-energy density $\mathcal{E}_B$. The competition between energy absorption and diffusion can be phenomenologically captured by (see details in the End Matter)
\begin{align}
\label{eq.rateequations_density}
\dot{\mathcal{E}}_{\mathrm{B}}&=\gamma_B,\
\dot{\mathcal{E}}_{\mathrm{D}}=\gamma_D-\Lambda R^{-1}\ ,
\end{align}
where we assume $R{\ll} L$,
$\gamma_{B/D}$ denotes the heating rate (shown in  Fig.~\ref{fig.rate} (a)) of the droplet and background at $\mathcal{E}_{B/D}$, while $\Lambda$ quantifies the cooling effect of energy diffusion through the droplet surface into the background. Linear response theory suggests that, $\Lambda{=}\kappa(\mathcal{E}_D{-}\mathcal{E}_B)$, with a constant $\kappa$.
We assume 
$\dot{R}{\sim} (\mathcal{E}_D{-}\mathcal{E}_B)/R$, which leads to diffusive growth if the energy difference is fixed. Yet, the concrete form of the droplet kinetic equation is not crucial for the physics discussed here,
as long as $\dot{R}$ is positive for $\mathcal{E}_D{>}\mathcal{E}_B$.  

Numerical integration of Eq.~\eqref{eq.rateequations_density} yields bistability, with two possible steady state solutions depending on the initial droplet size $R_i$. For a sufficiently large $R_i$, the droplet keeps growing as energy absorption through the bulk outweighs dissipation through the surface; by contrast, for small $R_i$ cooling effects dominate and the droplet equilibrates to the background energy with $R(t)$ converging to a finite value. {See details in End Matter and Sec.~3.3 in the SM~\cite{SM}.}

{{\it Spontaneous nucleation.--}
{Unlike the nucleation process with a `seed' droplet},
the energy density $\mathcal{E}_D$ of a spontaneous nucleation process is {\it a priori} unknown. 
We now show that the presence of spatial fluctuations in local energy density tends to spontaneously nucleate a critical droplet of an {\it optimal droplet energy density} from the low-energy background. This analysis also explains why superheating only arises for sufficiently low $\mathcal{E}_B$ and is absent in the high-frequency regime (see Fig.~\ref{fig.rate}(c)).}

From Eq.~\eqref{eq.rateequations_density} we find that the droplet grows if $R{>} R_c$, with
$R_c{=}{\kappa(\mathcal{E}_D{-}\mathcal{E}_B)}/({\gamma_D{-}\gamma_B}).$
The probability of a region with radius $R_c$
having a spontaneous fluctuation in energy density of magnitude ${\mathcal{E}}_{\mathrm{D}}{-}{\mathcal{E}}_{\mathrm{B}}$ can be estimated \begin{equation}\label{eq.probRc}
    P(R_c) {\sim} \exp\left[\frac{-(\mathcal{E}_D{-}\mathcal{E}_B)^2R_c^2}{2\sigma^2}\right]{=}\exp\left[\frac{-\kappa^2F(\mathcal{E}_D)}{2\sigma^2}\right]\ ,
\end{equation}
where $F(\mathcal{E}_D) {=} {(\mathcal{E}_D{-}\mathcal{E}_B)^{4}}/{(\gamma_D{-}\gamma_B)^2}$ and the fluctuation $\sigma$ depends on the background energy and system's specific heat, see details in End Matter.
Minimization of $F(\mathcal{E}_D)$ leads to the most probable droplet energy density, $\mathcal{E}_D^*$, by which one can further determine the corresponding droplet size. 
For concreteness, in Fig.~\ref{fig.rate}(b) we depict $F(\mathcal{E}_D)$ using the measured heating rate (Fig.~\ref{fig.rate}(a)) for a sufficiently low value of $\mathcal{E}_B$. 
Clearly, a nontrivial optimal solution $\mathcal{E}_D^*$ (red) appears that differs from $\mathcal{E}_B$ (blue). 
However, for higher background energy, as shown in Fig.~\ref{fig.rate}(c), $\mathcal{E}_D^*=\mathcal{E}_B$ and hence no spontaneous nucleation appears. This analysis thus justifies the numerical observation in Fig.~\ref{fig.phase}(c) where Floquet superheating only arises for sufficiently low energy initial states. 

In the high-frequency regime, $\gamma$ is generally exponentially suppressed as $\exp ({-}\mathcal{O}(\omega))$, resulting in a value of $R_c$ that is exponentially large in $\omega$. 
The corresponding nucleation probability, Eq.~\eqref{eq.probRc}, becomes doubly exponentially suppressed, $P(R_c){\sim} \exp[\exp({-}\mathcal{O}(\omega))]$, which is much smaller than the uniform heating rate. As a result, conventional homogeneous heating dominates.

{\it Discussion.--}
Floquet superheating should generally exist well beyond the particular setup considered here, as long as the heating rate can be sensitively suppressed for low-energy states. It thus presents another entry in our systematic understanding of possible heating pathways in driven systems. In our setup, this is achieved by SSSE, where interactions effectively echo out energy absorption channels, thereby providing a new avenue for stabilizing nonequilibrium phases of matter in driven systems. Generalizing this to {other states, }where quantum scars and spin helices may be employed to suppress homogeneous heating~\cite{sala2020ergodicity,surace2023weak,bhowmick2025granovskii,pizzi2025genuine,petrova2025finding,Theory2025Alessio}, is worth pursuing.

For numerical efficiency, we focus on classical systems, but we anticipate a similar phenomenon in quantum many-body systems. It remains an interesting open question to further explore the role of quantum fluctuations and {entanglement generation} during Floquet superheating as well as in droplet growth dynamics. As the large-scale simulation of quantum many-body systems poses formidable challenges for classical algorithms~\cite{krinitsin2025roughening,staszewski2025krylov}, controllable quantum simulators would be a natural platform to demonstrate Floquet superheating in practice
~\cite{rubio2020floquet,mi2022time,cheng2024observation,liu2025prethermalization}.

{Beyond fundamental interest, the remarkable sensitivity of the heating rate against local energy fluctuation may be employed to enhance the performance of magnetometry~\cite{moon2024discrete,lu2025dynamical}.}

\textit{Acknowledgments.---}
We thank Jianan Wang for stimulating discussions. This work is supported by Quantum Science and Technology-National Science and Technology Major Project
(No. 2024ZD0301800) and by the National Natural Science
Foundation of China (Grant No. 12474214), and by ``The Fundamental Research Funds for the Central Universities, Peking University”, and by ``High-performance Computing Platform of Peking University".
AP acknowledges support by Trinity College Cambridge.
HKJ acknowledges the support from the start-up funding from ShanghaiTech University.
This research was supported in part by grant NSF PHY-2309135 to the Kavli Institute for Theoretical Physics (KITP), {as well as by the Deutsche Forschungsgemeinschaft under grants FOR 5522 (project-id 499180199) and the cluster of excellence ct.qmat (EXC 2147, project-id 390858490).}
\bibliography{ref}

\part*{\scalebox{0.6}{End Matter}}
\begin{figure}[b]
    \centering
\includegraphics[width=0.7\linewidth]{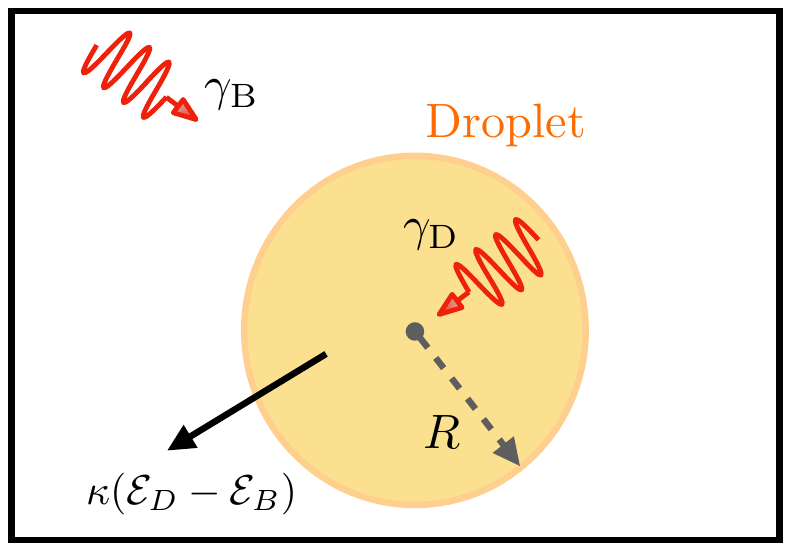}
\caption{Energy input and output for a droplet. $\gamma_{B}$ and $\gamma_{D}$  correspond to the energy absorption rate from the drive at the corresponding energy density, $\mathcal{E}_B$ and $\mathcal{E}_D$, respectively. $\kappa(\mathcal{E}_D-\mathcal{E}_B)$ quantifies the energy diffusion through the droplet surface to the background. 
}\label{fig.schematic}
\end{figure}

\addcontentsline{toc}{part}{End Matter}
\textit{Rate Equation.---} 
We now provide a detailed discussion of the rate equation that captures the competition between energy absorption and diffusion for a seeded droplet.
The total energy of the system, measured with respect to the ground state energy, $E_{\mathrm{tot}}$, follows 
$
\dot{E}_{\mathrm{tot}}=\dot{E}_{\mathrm{B}}+\dot{E}_{\mathrm{D}},
$
where two contributions $E_D$ and $E_B$ denote the total energy of the droplet and the rest of the system; see the schematic in Fig.~\ref{fig.schematic}. They follow the rate equations
\begin{equation}
    \label{eq.rateequations}
\dot{E}_{\mathrm{B}}=\gamma_BL^d+\Lambda R^{d-1}\ , \ \dot{E}_{\mathrm{D}}=\gamma_DR^d-\Lambda R^{d-1}\ ,
\end{equation}
where $\gamma_{B/D}$ corresponds to the energy density absorption rate at the corresponding energy density (red in Fig.~\ref{fig.schematic}) and $\Lambda$ quantifies the energy transport through the droplet surface to the background (black arrow in Fig.~\ref{fig.schematic}). 

For a sufficiently large system ($L{\gg} R$), we assume that the energy flowing out of the droplet quickly redistributes uniformly in the background. In practice, as long as the time scale for redistributing the energy is much shorter than other time scales of interest, this assumption is valid. We assume that the droplet keeps a sharp boundary during the time evolution for simplicity. 
Hence, one can rewrite Eq.~\eqref{eq.rateequations} using the energy density, 
\begin{align}
\label{eq.rate_SM}
\dot{\mathcal{E}}_{\mathrm{B}}&=\gamma_B\ ,\
\dot{\mathcal{E}}_{\mathrm{D}}=\gamma_D-\kappa R^{-1}(\mathcal{E}_D-\mathcal{E}_{B})\ ,
\end{align}
where we use $L{\gg} R$, the same as Eq.~\eqref{eq.rateequations_density} in the main text, and also we use $\Lambda=\kappa(\mathcal{E}_D-\mathcal{E}_{B})$ following the linear response theory. The droplet also grow as long as $\mathcal{E}_D>\mathcal{E}_B$, and we assume that it grows diffusively
\begin{equation}
\label{eq.R_growth}
    \dot{R}=\xi (\mathcal{E}_D-\mathcal{E}_B)/R\ ,
\end{equation}
where $\xi$ denotes the diffusion coefficient ~\cite{perez2008implementation}.  Note that we use these rate equations to analyze the growth of the droplet at early times. At long times, Floquet many-body systems eventually heat up to infinite temperature, and this simple rate equation may not be sufficient to capture the complex late-time evolution.

\begin{figure}
    \centering
    \includegraphics[width=0.9\linewidth]{ 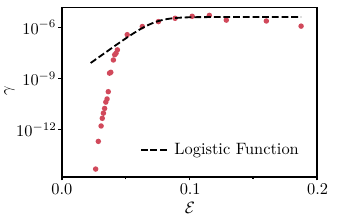}
    \caption{Heating rate $\gamma$ versus initial energy. We fit the curve using the logistic function (black dashed line), which works well for finite energies but fails to describe the sharp drop in the low-energy regime.}
    \label{fig.rate_fit}
\end{figure}

The following ansatz can capture the energy density dependence in the heating rate
\begin{equation}
\label{eq.heatingrate_function}
\gamma(\mathcal{E})=\frac{\gamma_{0}}{1+\exp{[-c(\mathcal{E}-\mathcal{E}^*)]}}\ ,
\end{equation}
with fitting parameters $\mathcal{E}^*$, $c$ and $\gamma_0$.
Hence, at high temperature ($\mathcal{E}^*\ll\mathcal{E}<0$), $\gamma$ converges to $\gamma_0$, without any energy dependence; while $\gamma$ is exponentially suppressed as $\mathcal{E}$ approaches the ground state energy. Such a crossover appears around the energy density $\mathcal{E}^*$ with a width determined by $1/c$. This ansatz works well at a finite energy (see black dashed line in Fig.~\ref{fig.rate_fit}) but cannot capture the sharp drop at low energy.

\textit{Spontaneous nucleation.---} Here we
show that the presence of spatial fluctuations in local
energy density tends to spontaneously nucleate a critical droplet.
For the droplet to heat, we require the rate of heat
input to exceed the rate of heat diffusion through the
surface.
For sufficiently low background energy density, we have $\gamma_B\ll\gamma_D$ as ${\mathcal{E}}_{\mathrm{B}}$ and ${\mathcal{E}}_{\mathrm{D}}$ are on opposite sides of the jump in $\gamma(\mathcal{E})$ as shown in Fig.~\ref{fig.rate}. For simplicity we can set  $\gamma_B=0$, so that the background does not heat at all. We generalize the discussion to $\gamma_B\neq0$ later. According to Eq.~\eqref{eq.rate_SM}, this leads to the requirement  
\begin{equation}
\label{eq.R_csimple}
R>R_c=\frac{\kappa({\mathcal{E}}_{\mathrm{D}}-{\mathcal{E}}_{\mathrm{B}})}{\gamma_D}\ .
\end{equation}
The probability of a region with radius $R_c$
having a spontaneous fluctuation in energy density of magnitude ${\mathcal{E}}_{\mathrm{D}}-{\mathcal{E}}_{\mathrm{B}}$ at temperature $T$  is given by
\begin{equation}
    P(R,{\mathcal{E}}_{\mathrm{D}}-{\mathcal{E}}_{\mathrm{B}})=
    \exp\left[
    -\frac{R^2({\mathcal{E}}_{\mathrm{D}}-{\mathcal{E}}_{\mathrm{B}})^2}{2k_B T^2 C}\ 
    \right]\ ,
    \label{eq:heat_ed_eb}
\end{equation}
with the specific heat $C$ and system's temperature $T$.
Now, we are of course not in thermal equilibrium, but we can borrow results from classical equipartition, which presumably applies to the effective Hamiltonian, to yield $C=k_B$ and, consequently, $\mathcal{E}=CT$, so that
\begin{equation}
    P(R_c)=\exp\left[
    -\frac{\kappa^2 ({\mathcal{E}}_{\mathrm{D}}-{\mathcal{E}}_{\mathrm{B}})^4}{2\gamma_D^2 {\mathcal{E}}_{\mathrm{B}}^2}
    \right] \ .
\end{equation}
This expression makes sense provided $\mathcal{E}_{B,D}$ are small and close to each other, and near the jump in $\gamma(\mathcal{E})$, as otherwise, the assumption of a single $T$ and describing the system and fluctuations around it does not apply. At any rate, Eq.~\eqref{eq:heat_ed_eb} is the slightly more general expression and may be more sensible to work with, and perhaps it is better to use 
\begin{equation}
\label{eq.p_RcT}     P(R_c)=\exp\left[
    -\frac{\kappa^2 ({\mathcal{E}}_{\mathrm{D}}-{\mathcal{E}}_{\mathrm{B}})^4}{2k_B\gamma_D^2 CT^2}
    \right] \ .
\end{equation}
More generally, we can also consider $\gamma_B\neq 0$, and the condition Eq.~\eqref{eq.R_csimple} can be modified as
\begin{equation}\label{eq.rc}
R_c=\frac{\kappa(\mathcal{E}_D-\mathcal{E}_B)}{\gamma_D-\gamma_B}\ .
\end{equation} 
According to the rate equation Eqs.~\eqref{eq.rate_SM} and \eqref{eq.R_growth}, this condition ensures that $\dot{\mathcal{E}}_D-\dot{\mathcal{E}}_B$ is positive for all times, and hence the rate of heat
input exceeds the rate of heat diffusion through the
surface. We also numerically confirm this in Sec.~3.3 in the SM~\cite{SM}. Consequently, Eq.~\eqref{eq.p_RcT} generalizes to
\begin{equation}
\label{eq.p_RcT_general}     P(R_c)=\exp\left[
    -\frac{\kappa^2 ({\mathcal{E}}_{\mathrm{D}}-{\mathcal{E}}_{\mathrm{B}})^4}{2(\gamma_D-\gamma_B)^2 CT^2k_B}
    \right] \ .
\end{equation}
In the main text, we define $\sigma^2=CT^2k_B$
and obtain Eq.~\eqref{eq.probRc}.

\clearpage

\let\addcontentsline\oldaddcontentsline
	\cleardoublepage
	\onecolumngrid
 \begin{center}
\textbf{\large{\textit{Supplementary Material} \\ \smallskip
	Floquet Superheating}}
    \\[1em]
Yang Hou$^{1,2}$,
Andrea Pizzi$^{3}$,
Huike Jin$^{4}$,
Johannes Knolle$^{5,6,7}$,
Roderich Moessner$^{2}$,
Hongzheng Zhao$^{1}$*
\\[.1cm]
{\small
$^{1}$\textit{\affA} \\
$^{2}$\textit{\affB} \\
$^{3}$\textit{\affC} \\
$^{4}$\textit{\affD} \\
$^{5}$\textit{\affE}\\
$^{6}$\textit{\affF} \\
$^{7}$\textit{\affG} \\
}
(Dated: \today)\\[1cm]
	\end{center}
	\renewcommand{\thefigure}{S\arabic{figure}}
        \setcounter{figure}{0}
    \renewcommand{\thesection}{SM\;\arabic{section}}
	\setcounter{section}{0}
	\renewcommand{\theequation}{S.\arabic{equation}}
        \setcounter{equation}{0}
    \renewcommand{\thesubsection}{\arabic{subsection}}
	\setcounter{section}{0}
    \tableofcontents
    \setcounter{page}{1}
\section{Further Discussions for Superheating Phenomenon}
Here, we present additional results that further illustrate the superheating phenomenon, including the effect of boundary conditions and the superheating behavior in other frequency windows as predicted by SSSE.
\subsection{Boundary Effects}
In this section, we discuss how boundary conditions affect the superheating phenomenon. 
\begin{figure}[h]
    \centering
\includegraphics[width=0.6\linewidth]{ 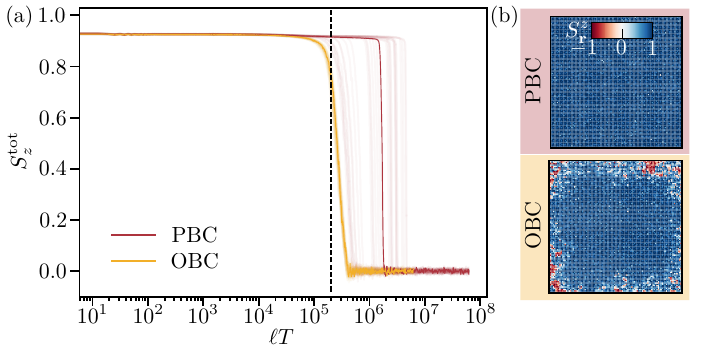}
\caption{(a) Dynamics of heating under OBC(yellow) and PBC(red) as reflected in the total $z-$magnetization. Systems with OBC clearly heat up faster than the PBC case. Each ensemble depicts 32 individual realizations, which stay close to each other under OBC, but not for PBC. We use $L=100$, $J=0.5$, $g=0.17$, $\delta\theta=0.12\pi$ for numerical simulations.
(b) Snapshots of the $z$-component spin configurations at the onset of heating under OBC (black dashed line in (a)) for PBC (up) and OBC (down). Heating sets in from the system's boundary under OBC. Meanwhile, the configuration remains unheated for PBC. The color denotes the value of the $z-$magnetization, such that blue and red colors correspond to spins pointing along the positive and negative $z$-directions, respectively. 
    }
    \label{fig.boundary}
\end{figure}

In the main text, we state that the SSSE is crucial for the appearance of the superheating phenomenon, which requires
\begin{equation}\label{eq.condition}
\omega_e=\frac{2J\bar{S}^z}{2n+1}\ ,\quad n\in \mathbb{Z}\ .
\end{equation}
It is important to notice that the coordination number for each site is contained in the prefactor. Therefore, the boundary condition, which directly determines the coordination number, plays an essential role in the superheating phenomenon. Concretely, under open boundary conditions (OBC), spins at edges and corners have different coordination numbers from those in the bulk, and hence SSSE no longer applies to the boundary. Consequently, thermalization sets in deterministically at the system's boundary. As shown in Fig.~\ref{fig.boundary}(a), from the same initial condition as the red curves in Fig~\ref{fig.dynamics}(a), OBC (yellow) leads to dynamics that is completely different from those under the periodic boundary condition (PBC, red). Under OBC, different trajectories remain close to each other, and their averaged heating time is much shorter than the PBC case. Microscopically, we demonstrate in Fig.~\ref{fig.boundary}(b) that under OBC, the onset of heating originates at the system boundaries, whereas the bulk remains nearly unheated. In contrast, at the same moment, evolution from the same initial state under PBC retains an ordered configuration throughout the system. 

\subsection{Superheating at $n=1$}\label{sec.n=1}
\begin{figure}[b]
    \centering
    \includegraphics[width=\linewidth]{ 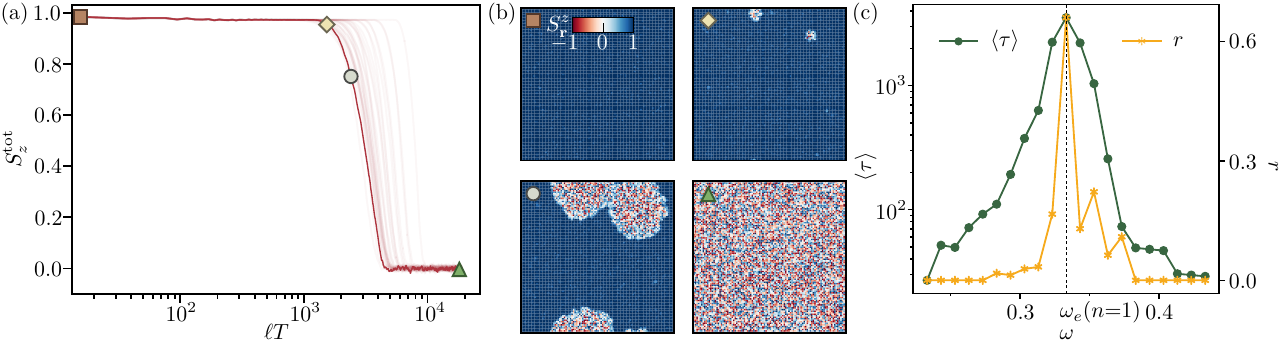}
\caption{Numerical evidence for superheating at $n=1$ (a) Dynamics of the total $z-$magnetization. 32 individual realizations that exhibit notable variance are depicted. 
(b) Snapshots of the $z$-component spin configurations. Floquet superheating occurs via the nucleation process. Here, we present one realization where two droplets nucleate, grow, merge, and eventually occupy the entire system. The color denotes the value of the $z-$magnetization, such that blue and red colors correspond to spins pointing along the positive and negative $z$-directions, respectively. (c) Mean value $\langle\tau\rangle$ and relative standard deviation $r$ of the prethermal lifetime. $\tau$ is extracted when $S_{z}^{\mathrm{tot}}$ first drops below 80 percent of its initial value. Both $\langle\tau\rangle$ and $r$ peaks at the echoing point $\omega_e(n=1)=1/3$. We use $\delta\theta=0.06\pi$ with $L=100$, $J=0.5$, $g=0.05$ for numerical simulations.}
    \label{fig.n=1}
\end{figure}
In the main text, we have demonstrated the superheating phenomenon for $n=0$, where the integer $n$ appears in the SSSE condition, Eq.~\eqref{eq.condition}. Here we show that similar behavior also appears for other integer values of $n$, e.g., $n=1$.

The dynamics of total magnetization is shown in Fig.~\ref{fig.n=1}(a), where heating occurs at strikingly distinct times in different realizations. As with $n=0$, the heating times follow a distribution with comparable mean and standard deviation. Microscopically, heating is also triggered by droplet nucleation. We highlight a particular trajectory where two droplets are nucleated out of the uniform background, as shown in Fig.~\ref{fig.n=1}(b). During the heating process, the two droplets first grow independently after nucleation (diamond) until they come into contact and merge (circle). 
For sufficiently large system sizes, one would expect the coexistence of a finite density of droplets in the system. 
For a fixed initial temperature, we depict the averaged prethermal lifetime $\langle\tau\rangle$ and the concomitant relative standard deviation $r$ around the neighborhood of the echoing point $\omega_e(n=1)$ in Fig.~\ref{fig.n=1}(c), where both of them peak at $\omega_e(n=1)$, and the maximum of $r$ takes a value of $\mathcal{O}(1)$. These typical features strongly support the presence of the superheating phenomenon at $n=1$.

The $n=1$ case, however, is more sensitive and fragile against perturbation compared with $n=0$. This is because $\omega_e$ becomes smaller as $n$ increases, pushing the system away from the high-frequency regime. Driven at such frequencies, the system undergoes rapid heating in the absence of SSSE. Accordingly, we need to choose a very small $\delta\theta$ for the initial spin configurations to prolong the heating time. In addition, the mean-field approximation in SSSE requires a small transverse field strength $g$ to ensure the alignment of spins along the $z$-axis. Since the rotation angle induced by the transverse field is $gT/4$, and the echoing driving period $T_e$ for $n=1$ is three times greater than that of $n=0$, $g$ must be reduced approximately by a factor of three to maintain a small rotation angle. Hence, a smaller transverse field is used here.

\section{Comparison of Conventional Spin Echo and State-selective Spin Echo}\label{sec.dtcssse}
In the main text, we emphasize that SSSE relies on the intrinsic many-body interaction and thereby exhibits a strong initial state dependence, making it fundamentally different from conventional spin echo, which is underpinned by global spin flips and shows no preference for initial states. To see this, we present a comparison between SSSE and the conventional spin echo, the latter of which leads to the prethermal discrete time crystal (DTC) order in 2D systems~\cite{pizzi2021classical,ye2021floquet}.
\begin{figure}[b]
    \centering
    \includegraphics[width=0.8\linewidth]{ 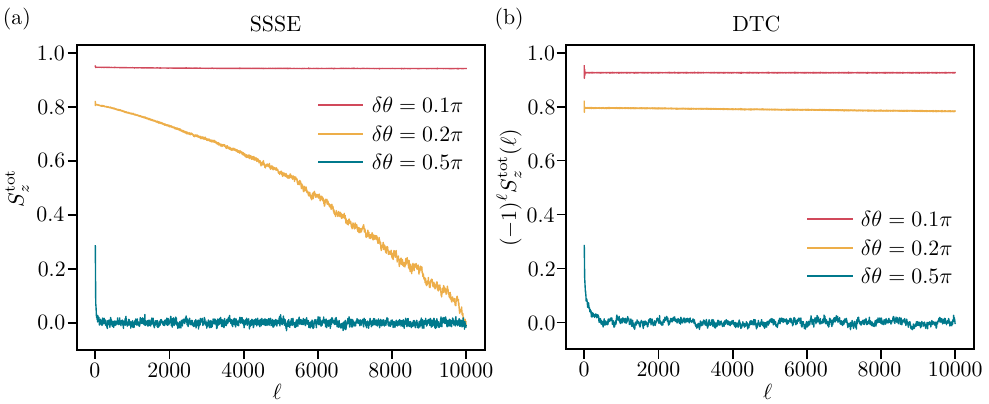}
    \caption{Comparison of conventional spin echo and SSSE. (a) Dynamics under the SSSE scheme, Eq.~\eqref{eq. Hamiltonian}. SSSE exhibits a strong dependence on the initial condition; only the low-energy state is prevented from heating. (b) Dynamics under the conventional spin echo (Eq.~\eqref{eq. dtcHamiltonian}), which leads to the prethermal DTC order. DTC order is insensitive to the initial condition and persists in the prethermal regime, as long as the initial state's temperature is below the critical temperature of the Ising phase transition. We use $L=100$, $J=0.5$, $g=0.17$ and $\phi_x=0.9\pi$ for numerical simulations. }
    \label{fig.dtc}
\end{figure}

To realize DTC, we consider the time-periodic Hamiltonian
\begin{equation}\label{eq. dtcHamiltonian}
    H(t){=}
    \begin{cases}
     H_1{\equiv} \frac{2\phi_x}{T} \sum_{\boldsymbol{r}} S_{\boldsymbol{r}}^x\quad\text{for}\ t\in \left[0,\frac{T}{2} \right)\\[8pt]
    H_2{\equiv} - \frac{J}{4}\sum_{{\boldsymbol{r}}{,}\boldsymbol{\delta}}S_{\boldsymbol{r}}^zS_
    {{\boldsymbol{r}}{+}\boldsymbol{\delta}}^z\quad \text{for}\ t\in \left[\frac{T}{2},{T}\right)
    \end{cases}\ ,
\end{equation}
with $T$ denoting the driving period. We remark that this protocol is equivalent to the one in the main text up to an additional global $x$-rotation. When we set $\phi_x=\pi$, i.e., the transverse field in the first step generates a $\pi$-pulse along the $x$ direction. The corresponding time evolution operator is a $\mathbb Z_2$ Ising symmetry generator, which commutes with the Ising interaction in the second step.
Therefore, after two cycles, the system is governed by the effective Hamiltonian that is precisely $H_2$. Consequently, the system exhibits period-doubling dynamics in the $z$-magnetization from a symmetry-broken initial state, leading to the DTC order. Crucially, a small deviation of $\phi_x$ away from $\pi$ does not destroy the DTC order immediately. Instead, it persists for a long time in the prethermal regime before the onset of notable heating~\cite{pizzi2021classical}. 

Fig.~\ref{fig.dtc}(a) depicts the dynamics of the total magnetization under the same driving protocol and driving parameters as those used in the main text. In contrast, Fig.~\ref{fig.dtc}(b) shows the dynamics of the DTC protocol, Eq.~\eqref{eq. dtcHamiltonian}, at $\phi_x=0.9\pi$, with all other parameters and initial states identical to those in panel (a). We take $(-1)^\ell S_z^{\mathrm{tot}}(\ell)$, with $\ell$ denoting the Floquet cycles, as the order parameter of DTC, which allows us to remove its oscillation between $\pm S_z$. For sufficiently low-energy initial states (red), $z$-magnetization remains close to their initial values in both cases. However, since SSSE is sensitive to the choice of the initial state, an increase in initial temperature (yellow) can quickly destabilize the system. On the contrary, conventional spin echo yields an emergent $\mathbb Z_2$ Ising symmetry in the effective Hamiltonian, and hence the DTC order generally persists provided the initial state's temperature is below the critical temperature of the thermal Ising phase transition. This comparison clearly reveals the fundamental difference between these two heating suppression mechanisms. Of course, for large enough initial randomness (blue), both systems heat up to the infinite-temperature state.

In summary, we highlight that SSSE is an interaction-induced, state-dependent heating suppression mechanism, which is fundamentally different from the conventional spin echo.

\section{Further Analysis for Droplet Theory}
Here, we provide further details on droplet theory. We present more numerical results to support the existence of the critical droplet size, discuss the scheme for fitting the heating rate, and numerically integrate the rate equations to obtain the bistable dynamics of the droplet.
\subsection{Critical Droplet Size}\label{sec.criticaldroplet}
The main text demonstrates that, a seeded droplet in the initial state generally accelerates heating. However, only when the initial droplet size is larger than a critical droplet size $R_c$, does spontaneous nucleation no longer appear. We now provide additional numerical evidence to extract $R_c$.
\begin{figure}[b]
    \centering
    \includegraphics[width=0.5\linewidth]{ 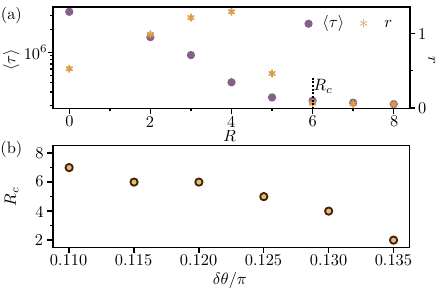}
    \caption{(a) Mean value $\langle\tau\rangle$ and relative standard deviation $r$ of the prethermal lifetime versus size of initially introduced droplet $R$ for $\delta\theta=0.12\pi$. As $R$ increases, $\langle\tau\rangle$ decreases and then saturates, and $r$ remains large before dropping to $0$. (b) Critical droplet size $R_c$ increases for weaker initial spatial randomness $\delta\theta$. We use $J=0.5$, $g=0.17$, and $L=100$ for numerical simulations.
    }
    \label{fig.droplet}
\end{figure}

In Fig.~\ref{fig.droplet}(a), we illustrate the mean and the relative standard deviation of perthermal lifetime versus various initial droplet sizes at $\omega=1$. As the droplet size increases, the averaged lifetime $\langle \tau \rangle$ decreases until it saturates to some value, whereas the deviation $r$ remains large over a range of droplet sizes before dropping to zero. Particularly, at $R=5$ where $\langle\tau\rangle$ appears to have converged but $r$ is appreciable, we observe that the heating times are narrowly distributed around the mean for most of the realizations, with a few exceptions significantly enlarging the deviation. Similar critical behavior also manifests near the phase boundary in Fig.~\ref{fig.phase}(c), and indicates that $r$ serves as a suitable indicator to determine the critical droplet size. In Fig.~\ref{fig.droplet}(b), we extract the critical droplet size $R_c$, defined as the minimal droplet size such that $r$ drops below 0.3, for different initial randomness $\delta\theta$. The critical droplet size tends to grow for smaller initial randomness. It happens because for smaller $\delta \theta$, the energy dissipation across the droplet's surface increases and hence one generally requires a larger critical droplet size to allow for more energy absorption through the droplet bulk. It should be noted that the critical sizes of seeded droplets generally do not coincide with those that spontaneously nucleate from the low-temperature background.

\subsection{Extraction of Heating Rate}\label{sec.ratefit}
\begin{figure}[h]
    \centering
    \includegraphics[width=0.5\linewidth]{ 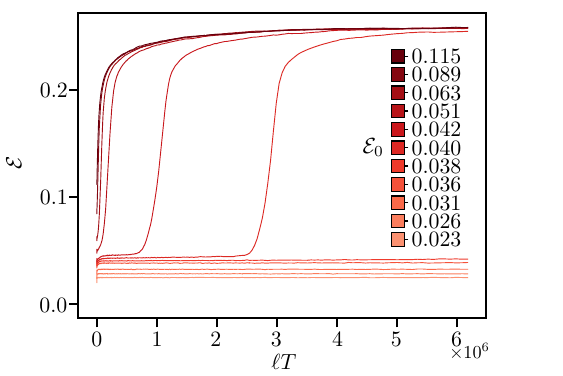}
    \caption{Traces of energy evolution. Low energy states ($\mathcal{E}_0\leq0.038$) do not heat up within the time window up to $\ell=10^6$ Floquet cycles. Initial states with higher energies ($\mathcal{E}_0=0.040,\ 0.042$) exhibit transient yet stable plateaus followed by fast heating. Further increasing initial energy leads to the conventional heating process. We use $L=150$, $J=0.5$, $g=0.05$ for numerical simulations.}
\label{fig.trace}
\end{figure}

In this section, we elaborate on how we extract the heating rate from the time evolution of energy for plotting Fig.~\ref{fig.rate}. Fig.~\ref{fig.trace} shows the dynamics of the energy density $\mathcal{E}$, measured with respect to the ground state energy, for single realizations, where darker-colored curves correspond to trajectories with larger initial energy densities. For a low-energy initial state, a short transient behavior is visible within the time window $t\le10^4$, after which the system evolves into a prethermal plateau. In Sec.~\ref{sec.prethermalproperty}, we will further discuss the correlation function of the system during this prethermal regime, but here we only focus on its heating behavior. 
The systems absorb energy at an approximately constant rate, which can be extracted by linearly fitting the energy time evolution after the early-time transient and before the $z$-magnetization drops to 95\% of their initial values. 
For low energy densities ($\mathcal{E}\leq0.038$), the heating rate is extremely small and the time traces are rather flat. Numerically, we find that the fitted heating rates can be even smaller than $10^{-11}$. This severely suppressed heating also leads to the sudden drop in the heating rate as seen in Fig.~\ref{fig.rate}
in the main text.

\subsection{Bistable Dynamics of the Rate Equations}
\label{sec.bistable}
In the main text, the following set of rate equations are used to describe the evolution of energy density and droplet size
\begin{equation}
    \dot{\mathcal{E}}_{\mathrm{B}}=\gamma_B,\
\dot{\mathcal{E}}_{\mathrm{D}}=\gamma_D-\kappa R^{-1}(\mathcal{E}_D-\mathcal{E}_{B})\ ,\ \dot{R}=\xi (\mathcal{E}_D-\mathcal{E}_B)/R\ .
\end{equation}
These equations capture the competition of energy absorption due to Floquet drive and the energy diffusion through the droplet surface. Therefore, these rate equations are bistable with two types of solutions: either the droplet keeps growing or the droplet cools down to the same energy density as the background. 
\begin{figure}[h]
    \centering
    \includegraphics[width=\linewidth]{ 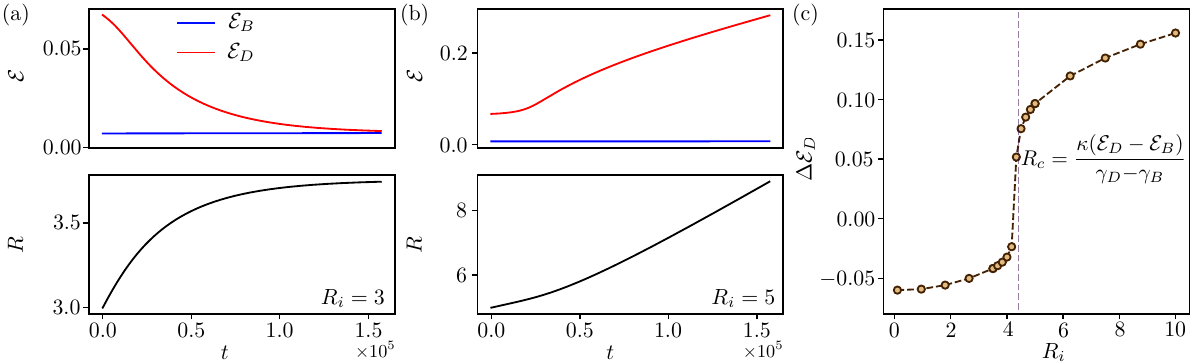}
    \caption{(a)\&(b) Bistable dynamics generated by the rate equations. (a) Initialized with droplet size $R_i=3$. The energy of the droplet diffuses to the background and the droplet size saturates near its initial value. (b) Initialized with droplet size $R_i=5$. The droplet continues to grow as it absorbs energy when the droplet size is sufficiently large.
    (c) Change in droplet energy, $\Delta\mathcal{E}_D=\mathcal{E}_D(t_f)-\mathcal{E}_D(0)$, after evolving to $t_f=6\times10^4$. A sharp transition signifies the critical droplet size, which is upper-bounded by the theoretical estimation, labeled by the purple dashed line. We take the initial energy as $\mathcal{E}_B(0)=0.007$, $\mathcal{E}_D(0)=0.067$, and use $\kappa=10^{-4}$, $\xi=10^{-3}$, and $\gamma(\mathcal{E})=\frac{\gamma_{0}}{1+\exp{[-c(\mathcal{E}-\mathcal{E}^*)]}}$ with $\gamma_0=4.18\times10^{-6}$, $c=123.01$, $\mathcal{E}^*=0.073$.}
    \label{fig.rateeq}
\end{figure}

To show this, we numerically integrated these equations, and the results are shown in Fig.~\ref{fig.rateeq}. In Fig.~\ref{fig.rateeq}(a)\&(b), we present two distinct dynamical behaviors initialized with different initial droplet sizes $R_i$. In panel (a), the system starts with a smaller $R_i$, and the energy diffusion through the surface beats the energy absorption in the bulk. Consequently, the droplet energy density cools down and the droplet size saturates near its initial value. Fig.~\ref{fig.rateeq}(b) shows the dynamics with an initial droplet that exceeds the critical droplet size, where the droplet keeps expanding while absorbing energy. The transition point between the two types of dynamics defines the critical droplet size. To extract the critical droplet size, we numerically simulate the rate equations up to $t_f=6\times10^4$, and plot the energy change of the droplet $\Delta\mathcal{E}_D=\mathcal{E}_D(t_f)-\mathcal{E}_D(0)$ as a function of initial droplet size $R_i$ in Fig.~\ref{fig.rateeq}(c). During evolution, a droplet that cools down exhibits a negative change in energy, while a continuously growing droplet undergoes an increase in energy. As shown in Fig.~\ref{fig.rateeq}(c), a sharp transition from negative to positive value occurs at $R_i\approx4.5$. 

Compared with our theoretical prediction of the critical droplet size (purple dashed line), cf.~Eq.~\eqref{eq.rc}, the rate equation yields a smaller but close value of the critical droplet size. In fact, the transition point determined by the rate equations is always upper-bounded by the theoretical estimation of $R_c$ (purple), and their deviation depends on the concrete value of the growth rate $\xi$ of the droplet dynamics. Larger $\xi$ facilitates droplet growth, yielding a smaller critical droplet size obtained by the rate equation. On the other hand, $\xi=0$ implies that the droplet growth rate is much smaller than any other energy scales in the system. In this case, one can obtain the analytical prediction for the critical droplet size Eq.~\eqref{eq.rc}. Based on our numerical simulation of the many-body dynamics, we estimate that $\xi\approx10^{-3}$, which we use in our numerical integration of the rate equation. 

In the numerical simulation of the rate equations, 
the heating rate $\gamma$ is modeled as the logistic function [cf.~Eq.~\eqref{eq.heatingrate_function}], using the same parameters as those in the fitted line in Fig.~\ref{fig.rate}. 
To estimate the diffusion constant $\kappa$, we first prepare half of the system at the infinite temperature and the other half at a low temperature. Then we quench the system with the averaged Hamiltonian using the same parameter as in the main text and track the energy transport across the interface. This leads to the approximation of  $\kappa\approx10^{-4}$. 

\section{Property of the Prethermal Plateau} 
\label{sec.prethermalproperty}
Now we provide a finer description of SSSE and reveal the appearance of an emergent antiferromagnetic (AFM) 
order in the $y$-direction during the prethermal regime. It relies on the existence of a set of exactly solvable periodic orbits induced by the driving protocol. Importantly, the appearance of the AFM order seems to be strongly correlated with the severely suppressed heating rate at low energy densities.

\subsection{Existence of Exact Periodic Orbits}\label{sec.closeorbit}
To show the existence of the periodic orbits, we begin with the Floquet operator for the quantum system and later discuss the classical system. The Floquet operator for our driving protocol reads
$U_F=e^{-iH_1\frac{T}{4}}e^{-iH_2\frac{T}{2}}e^{-iH_1\frac{T}{4}}$. The operator can be rewritten as
\begin{equation}
    \begin{aligned}
    U_F&=e^{-iH_1\frac{T}{4}}e^{-iH_2\frac{T}{2}}e^{+i\pi\sum_{\boldsymbol{r}}S_{\boldsymbol{r}}^z}{e^{-i\pi\sum_{\boldsymbol{r}}S_{\boldsymbol{r}}^z}e^{-iH_1\frac{T}{4}}e^{+i\pi\sum_{\boldsymbol{r}}S_{\boldsymbol{r}}^z}}e^{-i\pi\sum_{\boldsymbol{r}}S_{\boldsymbol{r}}^z}\\
    &={e^{-iH_1\frac{T}{4}}e^{-iH_2\frac{T}{2}+i\pi\sum_{\boldsymbol{r}}S_{\boldsymbol{r}}^z}{e^{+iH_1\frac{T}{4}}}}e^{-i\pi\sum_{\boldsymbol{r}}S_{\boldsymbol{r}}^z}\\
    &={e^{i\frac{T}{2}\frac{J}{4}\sum_{{\boldsymbol{r}}{,}\boldsymbol{\delta}}\widetilde{S}_{\boldsymbol{r}}^z\widetilde{S}_
    {{\boldsymbol{r}}{+}\boldsymbol{\delta}}^z+i\pi\sum_{\mathbf{r}}\widetilde{S}_{\mathbf{r}}^z}}e^{-i\pi\sum_{\mathbf{r}}S_{\mathbf{r}}^z}
\end{aligned}\ ,
\end{equation}
where $\widetilde{S}_{\boldsymbol{r}}^z={S}_{\boldsymbol{r}}^z\cos{\frac{gT}{4}}+{S}_{\boldsymbol{r}}^y\sin{\frac{gT}{4}}$. Consider two Floquet cycles, we have 
\begin{equation}
    \begin{aligned}
        U_F^2&={e^{i\frac{T}{2}\frac{J}{4}\sum_{{\boldsymbol{r}}{,}\boldsymbol{\delta}}\widetilde{S}_{\boldsymbol{r}}^z\widetilde{S}_
    {{\boldsymbol{r}}{+}\boldsymbol{\delta}}^z+i\pi\sum_{\mathbf{r}}\widetilde{S}_{\mathbf{r}}^z}}{e^{-i\pi\sum_{\mathbf{r}}S_{\mathbf{r}}^z}{e^{i\frac{T}{2}\frac{J}{4}\sum_{{\boldsymbol{r}}{,}\boldsymbol{\delta}}\widetilde{S}_{\boldsymbol{r}}^z\widetilde{S}_
    {{\boldsymbol{r}}{+}\boldsymbol{\delta}}^z+i\pi\sum_{\mathbf{r}}\widetilde{S}_{\mathbf{r}}^z}}e^{-i\pi\sum_{\mathbf{r}}S_{\mathbf{r}}^z}}\\
    &={e^{i\frac{T}{2}\frac{J}{4}\sum_{{\boldsymbol{r}}{,}\boldsymbol{\delta}}\widetilde{S}_{\boldsymbol{r}}^z\widetilde{S}_
    {{\boldsymbol{r}}{+}\boldsymbol{\delta}}^z+i\pi\sum_{\mathbf{r}}\widetilde{S}_{\mathbf{r}}^z}}{e^{i\frac{T}{2}\frac{J}{4}\sum_{{\boldsymbol{r}}{,}\boldsymbol{\delta}}\widetilde{P}_{\boldsymbol{r}}^z\widetilde{P}_
    {{\boldsymbol{r}}{+}\boldsymbol{\delta}}^z+i\pi\sum_{\mathbf{r}}\widetilde{P}_{\mathbf{r}}^z}}
    \end{aligned}\ ,
\end{equation}
where $\widetilde{P}_{\boldsymbol{r}}^z={S}_{\boldsymbol{r}}^z\cos{\frac{gT}{4}}-{S}_{\boldsymbol{r}}^y\sin{\frac{gT}{4}}$. The equation above suggests that the evolution over two driving periods involves an alternation of two Ising models with longitudinal fields, with the Ising axis being $\widetilde{P}_{\boldsymbol{r}}^z$ and $\widetilde{S}_{\boldsymbol{r}}^z$. Note, the derivation for the Floquet operator in a classical system is the same as in the quantum system, see details in Ref.~\cite{mori2018floquet}. Crucially, this type of evolution enables us to uncover exactly solvable closed orbits in classical systems. To see this, we first divide the square lattice into two sublattices $A$ and $B$, consisting of alternating sites. Then, we initialize the system with translation-symmetric sublattices by setting $\boldsymbol{S}_A=\boldsymbol{S}_\pm$, $\boldsymbol{S}_B=\boldsymbol{S}_\mp$, where $\boldsymbol{S}_\pm=(0,\pm\sin{\frac{gT}{4}}\cos{\frac{gT}{4}})$, pointing along the Ising axes $\widetilde{P}_{\boldsymbol{r}}^z$ and $\widetilde{S}_{\boldsymbol{r}}^z$. One can show that this configuration stays unchanged under the evolution of both longitudinal-field Ising models. For the first operator, $\boldsymbol{S}_-$ stays unchanged, while $\boldsymbol{S}_+$ rotates around $\boldsymbol{S}_-$ for $2\pi$. For the second operator, $\boldsymbol{S}_-$ rotates for $2\pi$ with static $\boldsymbol{S}_+$. The $2\pi$ rotations, which correspond to the effect of applying SSSE twice, give rise to closed orbits in phase space.

We then perform the linear stability analysis of these closed orbits. We slightly perturb these orbits as $\boldsymbol{S}_{\boldsymbol{r}\in A}=\boldsymbol{S}_++\boldsymbol{\epsilon}_{\boldsymbol{r}}$, $\boldsymbol{S}_{\boldsymbol{r}\in B}=\boldsymbol{S}_-+\boldsymbol{\epsilon}_{\boldsymbol{r}}$, where $\boldsymbol{\epsilon}_{\boldsymbol{r}}$ is small perturbation with $\epsilon_{\boldsymbol{r}}\ll1$. Substituting this state into the evolution maps and expanding up to the linear order of perturbations, we have
\begin{equation}
    \boldsymbol{\epsilon}_{\boldsymbol{r}}(2T)
    =\left\{
    \begin{aligned}
        &\mathbf{M}\boldsymbol{\epsilon}_{\boldsymbol{r}}(0)+\mathbf{N}_\mathrm{A}\sum_{\boldsymbol{\delta}}\boldsymbol{\epsilon}_{\boldsymbol{r+\delta}}(0)\quad\mathrm{for}\ \boldsymbol{r}\in A\\[8pt]
        &\mathbf{M}'\boldsymbol{\epsilon}_{\boldsymbol{r}}(0)+\mathbf{N}_\mathrm{B}\sum_{\boldsymbol{\delta}}\boldsymbol{\epsilon}_{\boldsymbol{r+\delta}}(0)\quad\mathrm{for}\ \boldsymbol{r}\in B
    \end{aligned}\right.\ ,
\end{equation}
where
\begin{equation}
    \mathbf{M}{=}\left(
\begin{array}{ccc}
 -\cos \left(\pi  \cos{\frac{g T}{2}}\right) & -\cos \frac{g T}{4} \sin \left(\pi  \cos{\frac{g T}{2}}\right) & -\sin \frac{g
   T}{4} \sin \left(\pi  \cos{\frac{g T}{2}}\right) \\
 \cos{\frac{g T}{4}}\sin \left(\pi  \cos{\frac{g T}{2}}\right) & \sin ^2{\frac{g T}{4}}{-}\cos ^2{\frac{g T}{4}}\cos \left(\pi 
   \cos{\frac{g T}{2}}\right) & -\sin \frac{g T}{2}\cos ^2\left(\frac{1}{2} \pi  \cos{\frac{g T}{2}}\right) \\
 \sin \frac{g T}{4} \sin \left(\pi  \cos \frac{g T}{2}\right) & -\sin \frac{g T}{2} \cos ^2\left(\frac{1}{2} \pi  \cos
   \frac{g T}{2}\right) & \cos ^2\frac{g T}{4}{-}\sin ^2\frac{g T}{4} \cos \left(\pi  \cos{\frac{g T}{2}}\right) \\
\end{array}
\right)\ ,
\end{equation}
\begin{equation}
    \mathbf{M}'{=}\left(
\begin{array}{ccc}
 -\cos \left(\pi  \cos{\frac{g T}{2}}\right) & -\cos \frac{g T}{4} \sin \left(\pi  \cos{\frac{g T}{2}}\right) & \sin \frac{g
   T}{4} \sin \left(\pi  \cos{\frac{g T}{2}}\right) \\
 \cos{\frac{g T}{4}}\sin \left(\pi  \cos{\frac{g T}{2}}\right) & \sin ^2{\frac{g T}{4}}{-}\cos ^2{\frac{g T}{4}}\cos \left(\pi 
   \cos{\frac{g T}{2}}\right) & \sin \frac{g T}{2}\cos ^2\left(\frac{1}{2} \pi  \cos{\frac{g T}{2}}\right) \\
 \sin \frac{g T}{4} \sin \left(\pi  \cos \frac{g T}{2}\right) & \sin \frac{g T}{2} \cos ^2\left(\frac{1}{2} \pi  \cos
   \frac{g T}{2}\right) & \cos ^2\frac{g T}{4}{-}\sin ^2\frac{g T}{4} \cos \left(\pi  \cos{\frac{g T}{2}}\right) \\
\end{array}
\right)\ ,
\end{equation}
\begin{equation}
    \mathbf{N}_\mathrm{A}=\left(
\begin{array}{ccc}
 0 & -\pi \sin \frac{g T}{4} \sin \frac{g T}{2} & -\pi \sin \frac{g T}{2} \cos \frac{g T}{4} \\
 0 & 0 & 0 \\
 0 & 0 & 0 \\
\end{array}
\right)\ ,
\end{equation}
\begin{equation}
    \mathbf{N}_\mathrm{B}=\left(
\begin{array}{ccc}
 0 & \pi \sin \frac{g T}{4} \sin \frac{g T}{2} \cos \left(\pi  \cos \frac{g T}{4}\right) & -\pi  \sin \frac{g
   T}{2} \cos \frac{g T}{4} \cos \left(\pi  \cos \frac{g T}{4}\right) \\
 0 & -\frac{1}{2} \pi  \sin ^2\frac{g T}{2} \sin \left(\pi  \cos \frac{g T}{4}\right) & \pi  \sin \frac{g T}{2} \cos
   ^2\frac{g T}{4} \sin \left(\pi  \cos \frac{g T}{4}\right)\\
 0 & \pi  \sin ^2\frac{g T}{4} \sin \frac{g T}{2} \sin \left(\pi  \cos \frac{g T}{4}\right) & -\frac{1}{2} \pi   \sin
   ^2\frac{g T}{2} \sin \left(\pi  \cos \frac{g T}{4}\right) \\
\end{array}
\right)\ .
\end{equation}
We perform a Fourier transformation 
\begin{equation}
    \boldsymbol{\epsilon}_{\boldsymbol{r}\in A/B}=\sum_{\boldsymbol{q}}\boldsymbol{\epsilon}_{\boldsymbol{q}}^{A/B}e^{i\boldsymbol{q}\cdot\boldsymbol{r}}\  ,
\end{equation}
and obtain
\begin{equation}
    \left(\begin{aligned}
    &\boldsymbol{\epsilon}_{\boldsymbol{q}}^{A}\\&\boldsymbol{\epsilon}_{\boldsymbol{q}}^{B}
    \end{aligned}\right)(2T)=\left(\begin{array}{cc}
    \mathbf{M} &\alpha_{\boldsymbol{q}}\mathbf{N}_\mathrm{A}\\[8pt]
    \alpha_{\boldsymbol{q}}\mathbf{N}_\mathrm{B}&\mathbf{M}'
    \end{array}\right) \left(\begin{aligned}
    &\boldsymbol{\epsilon}_{\boldsymbol{q}}^{A}\\&\boldsymbol{\epsilon}_{\boldsymbol{q}}^{B}
    \end{aligned}\right)(0)\ ,
\end{equation}
where $\alpha_{\boldsymbol{q}}=2(\cos q_x+\cos q_y)$. According to the linear stability theory, we have
\begin{equation}
    \lambda\left(\begin{aligned}
    &\boldsymbol{\epsilon}_{\boldsymbol{q}}^{A}\\&\boldsymbol{\epsilon}_{\boldsymbol{q}}^{B}
    \end{aligned}\right)=\frac{d}{dt}\left(\begin{aligned}
    &\boldsymbol{\epsilon}_{\boldsymbol{q}}^{A}\\&\boldsymbol{\epsilon}_{\boldsymbol{q}}^{B}
    \end{aligned}\right)\approx\frac{1}{2T}\left[\left(\begin{aligned}
    &\boldsymbol{\epsilon}_{\boldsymbol{q}}^{A}\\&\boldsymbol{\epsilon}_{\boldsymbol{q}}^{B}
    \end{aligned}\right)(2T)-\left(\begin{aligned}
    &\boldsymbol{\epsilon}_{\boldsymbol{q}}^{A}\\&\boldsymbol{\epsilon}_{\boldsymbol{q}}^{B}
    \end{aligned}\right)(0)\right]=\left(\begin{array}{cc}
    \mathbf{M}-\mathbb{I} &\alpha_{\boldsymbol{q}}\mathbf{N}_\mathrm{A}\\[8pt]
    \alpha_{\boldsymbol{q}}\mathbf{N}_\mathrm{B}&\mathbf{M}'-\mathbb{I}
    \end{array}\right) \left(\begin{aligned}
    &\boldsymbol{\epsilon}_{\boldsymbol{q}}^{A}\\&\boldsymbol{\epsilon}_{\boldsymbol{q}}^{B}
    \end{aligned}\right)(0)\ .
\end{equation}
Solving the eigenproblem, we get the eigenvalues $\lambda=\{0,\ {i\frac{\pi}{2T}(1\pm\cos\frac{gT}{2})}\}$, with each of them two-fold degenerate. The Lyapunov exponent for this system is $\lambda_L\coloneq\max\{\mathrm{Re}(\lambda)\}=0$, suggesting the closed orbits are stable at the leading order of the perturbative expansion.
Interestingly, {$\lambda$ is independent of the momentum $\boldsymbol q$ because of the special form of the matrix $\mathbf N_{A/B}$}.
In many-body systems, higher-order non-linear effects will appear and hence these periodic orbits eventually become unstable with a finite lifetime. However, as shown in the next section, for sufficiently low-energy initial states, these periodic orbits lead to an emergent AFM order in the $y$ direction of the magnetization, which persists throughout the metastable prethermal plateau before the onset of notable heating. 
\subsection{Correlation Function of the Prethermal Plateau}\label{sec.corr}
\begin{figure}[t]
    \centering
    \includegraphics[width=0.8\linewidth]{ 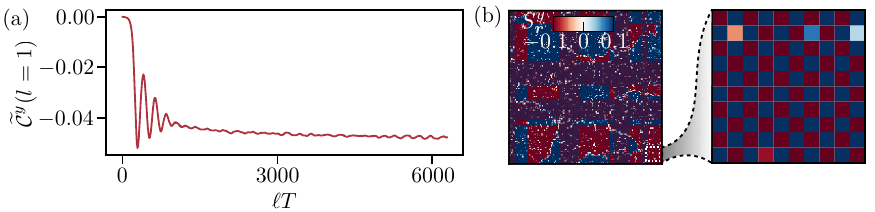}
    \caption{(a) Dynamics of correlation function of $y$-component at $l=1$, $\widetilde{\mathcal C}^y(l=1)$, showing the formation of the AFM order from an uncorrelated initial state. (b) Real-space spin configuration of the $y$ component to demonstrate the AFM order.}
    \label{fig.afmorder}
\end{figure}
\begin{figure}[b] 
    \centering
    \includegraphics[width=0.8\linewidth]{ 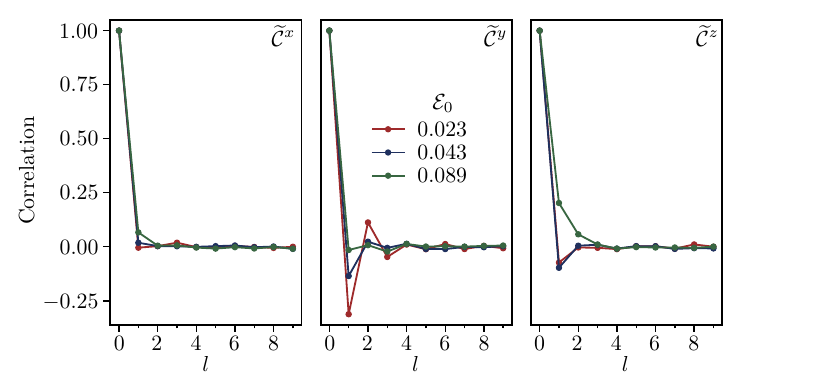}
    \caption{Normalized connected correlation function $\widetilde{\mathcal C}^\alpha(l)$ versus spatial distance $l$ in the prethermal regime. Left: correlation function of the $x$ component of spins, which is non-zero only at $l=0$. Middle: correlation function of the $y$ component and a negative value at $l=1$ suggests AFM order for sufficiently low initial energies (blue and red). Right: correlation function of the $z$ component of spins where weak AFM order also appears when the initial energy is low (blue and red). FM order is established as the initial energy increases, corresponding to prethermal order with respect to the averaged Ising Hamiltonian.}
    \label{fig.corr_com}
\end{figure}
In Fig.~\ref{fig.trace}, the energy traces exhibit a short-time transient period before entering  a long-lived prethermal plateau, during which an emergent AFM order in the $y$-magnetization gradually develops. To see this, we consider the spatial spin-spin correlation function
\begin{equation}
\mathcal{C}^\alpha(l)=\left\langle\sum_{r_y}\left(S^\alpha_{(0,r_y)}-\bar{S}^\alpha\right)\left(S^\alpha_{(l,r_y)}-\bar{S}^\alpha\right)\right\rangle=\left\langle\sum_{r_y}\delta S^\alpha_{{(0,r_y)}}\cdot\delta S^\alpha_{{(l,r_y)}}\right\rangle\ ,
\end{equation}
where $\alpha\in\{x,y,z\}$ is the components of the spin. In particular, we will normalize the correlation function as $\widetilde{\mathcal C}^\alpha(l)=\mathcal{C}^\alpha(l)/\mathcal{C}^\alpha(0)$, which allows us to compare correlations for different components $\alpha$. 

To show the spontaneous formation of this AFM order in our system, we first prepare a low-temperature system with $\delta\theta=0.01\pi$, with no order in y-magnetization, and monitor the evolution of the correlation function $\widetilde{\mathcal C}^y(l=1)$. As shown in Fig.~\ref{fig.afmorder}(a), at the early stage, the correlation function starts from 0 and saturates at a negative value, suggesting that the system establishes an AFM order as it settles into the prethermal regime. In Fig.~\ref{fig.afmorder}(b), we plot the real-space spin configuration of the $y$ component at the end of the dynamics in panel (a), where the AFM pattern is clearly visible. 

We further investigate the correlation function $\widetilde{\mathcal C}^\alpha(l)=\mathcal{C}^\alpha(l)/\mathcal{C}^\alpha(0)$ along different directions and at various distances after the system saturates to the prethermal regime. In Fig.~\ref{fig.corr_com}, we present the correlation function for three initial energies. In the left panel, $\widetilde{\mathcal C}^x(l)$ is clearly nonzero only at $l=0$, suggesting the absence of spatial correlations in $x$-direction. The middle panel shows the correlation in the $y$-component. When the initial energy is high (green), there is no spatial correlation. For sufficiently low-energy density initial states (blue and red), however, the correlation function becomes negative for odd $l$ and positive for even ones, indicating the development of the AFM order along the $y$ direction. Similarly, for sufficiently low-energy initial state, the $z$-component also exhibits weak AFM order (blue and red) in the right panel. For higher initial energies, FM order appears in the $z$-direction, corresponding to the prethermal order with respect to the averaged Hamiltonian $H_{\mathrm{eff}}$. 

\begin{figure}[h]
    \centering
    \includegraphics[width=0.5\linewidth]{ 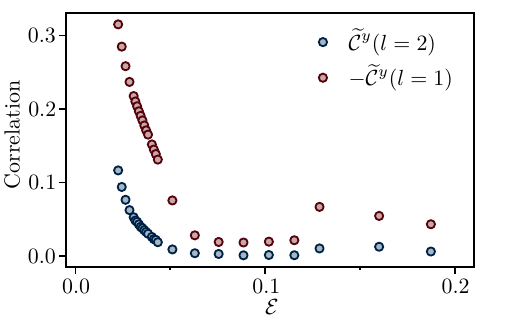}
    \caption{Spatial correlation in $y$ direction of nearest and next-nearest neighbors as a function of initial energy $\mathcal{E}_0$. Both correlation functions at $l=1$ and $l=2$ exhibit non-zero values for low initial energies, while they converge to small values close to 0 as energy increases.}
    \label{fig.corr_12}
\end{figure}

Moreover, we analyze the dependence of the correlation function $\widetilde{\mathcal C}^y(l)$ at $l=1$ and $l=2$ on the initial energies. As shown in Fig.~\ref{fig.corr_12},  both of them exhibit notable non-zero values at low energies and converge to $0$ as the initial energy increases. This suggests the robustness of the AFM order against small but finite initial thermal fluctuations. 

Crucially, we also find that the heating rate shown in Fig.~\ref{fig.rate} strongly correlates with the AFM order during the prethermal plateau, and more precisely we find $\gamma(\mathcal{E}) \sim \exp[-\widetilde{\mathcal C}^y(\mathcal{E})]$. Justifying this relation goes beyond the scope of this work, and we leave it for future investigations.
\end{document}